\documentclass[prd, twocolumn, nofootinbib]{revtex4}

\usepackage{epsfig} 

\newcommand{\beq}{\begin{equation}}
\newcommand{\eeq}{\end{equation}}
\newcommand{\beqa}{\begin{eqnarray}}
\newcommand{\eeqa}{\end{eqnarray}}
\newcommand{\om}{\Omega_m}

\def\fun#1#2{\lower3.6pt\vbox{\baselineskip0pt\lineskip.9pt
  \ialign{$\mathsurround=0pt#1\hfil##\hfil$\crcr#2\crcr\sim\crcr}}}

\begin{document} 

\title{Importance of Supernovae at $z<0.1$ for Probing Dark Energy} 
\author{Eric V.\ Linder} 
\affiliation{Berkeley Lab, University of California, Berkeley, CA 94720} 

\date{\today} 

\begin{abstract} 
Supernova experiments to 
characterize dark energy require a well designed low redshift program; 
we consider this for both ongoing/near term (e.g.\ Supernova Legacy Survey) 
and comprehensive future (e.g.\ SNAP) experiments. 
The derived criteria are: a supernova sample centered near $z\approx0.05$ 
comprising 150-500 (in the former case) and 300-900 (in the latter case) 
well measured supernovae. 
Low redshift Type Ia supernovae play two important roles for cosmological 
use of the supernova distance-redshift relation: as an anchor for the 
Hubble diagram and as an indicator of possible systematics.  An innate 
degeneracy in cosmological distances implies that 300 
nearby supernovae nearly saturate their cosmological leverage for the 
first use, and their optimum central redshift is $z\approx0.05$.  This 
conclusion is strengthened upon including velocity flow and magnitude 
offset systematics.  Limiting 
cosmological parameter bias due to supernova population drift (evolution) 
systematics plausibly increases the requirement for the second use to 
less than about 900 supernovae.  

\end{abstract} 

\maketitle 

\section{Introduction} \label{sec:intro} 

Type Ia supernovae (SN) observations discovered the acceleration of the 
universe \cite{perl99,riess98} and have proved central in progress elucidating 
the nature of the dark energy responsible \cite{snls,knop03,riess04}. 
SN provide a clear, direct, and mature method for mapping the expansion 
history of the universe, $a(t)$, with their measured flux giving the distance 
and hence lookback time $t$ through the cosmological inverse square law and 
their redshift $z$ giving the scale factor $a$. 

Ground-based SN surveys are already underway to obtain hundreds of SN 
in the range $z=0.2-0.9$, suitable for measuring an averaged, or 
assumed constant, dark energy equation of state. 
Plans are well advanced for a comprehensive SN experiment aimed at 
accurate determination of dark energy properties including dynamics, 
in the form of a 
wide field space telescope exquisitely characterizing some 2000 SN 
over the range $z=0.2-1.7$, with launch planned 
by 2013.  We consider here what specific low redshift 
(``local'': $z\lesssim0.2$) SN program would strengthen either of 
the higher redshift 
SN programs, combined with Planck CMB constraints on the distance to last 
scattering.  (We do not consider adding further probes, since it is useful 
to obtain a answer through purely geometric probes, as well as comparing 
results from different techniques.) 

Just as \cite{lh03,fhlt} brought into sharp relief the importance of SN at 
$z\gtrsim1.5$ for probing dark energy, due to their ability to break 
degeneracies, control systematics, and realistically follow dark energy 
dynamics, here we show similar crucial roles of SN at $z\lesssim0.1$. 
Local SN serve two key purposes: leverage in breaking degeneracies by 
anchoring the low redshift Hubble, or magnitude-redshift, diagram, 
and providing a well characterized 
set of SN to search for systematic magnitude effects through 
subclassification (``like subsets'').  In \S\ref{sec:cos} we demonstrate 
the cosmological need for a low redshift sample, and follow this in 
\S\ref{sec:nz} with detailed analysis of the numbers and redshift 
distribution for optimum complementarity with the higher redshift SN. 
Issues of systematics including evolution and intrinsic dispersion 
are discussed in \S\ref{sec:sys}, and a two stage survey program 
outlined to fulfill the required science criteria.

\section{Anchoring the Hubble Diagram} \label{sec:cos} 

The Hubble diagram plots the calibrated peak magnitude $m$, or flux received, 
of SN vs.\ their redshift.  Since the magnitude is a convolution of the 
intrinsic luminosity and the distance, to employ the SN data as a 
distance probe from which we extract cosmological parameters we need to 
anchor the diagram at low redshift where the distance becomes independent 
of cosmology.  This serves to constrain the intrinsic luminosity (combined 
with the Hubble constant) in a nuisance parameter ${\cal M}$. 

Also, at low redshift the cosmological information enters first in the 
form of the 
deceleration parameter $q_0=(1+3w_0\Omega_w)/2$, where $\Omega_w$ is the 
dark energy density in units of the critical density and $w_0$ its 
present equation of state ratio.  If we concentrate on 
the dark energy equation of state plane $w_0$-$w_a$ where $w(a)=w_0+w_a(1-a)$, 
then at low redshift there is no dependence on $w_a$ and confidence contours 
from local SN data would be vertical.  As we consider SN at higher redshifts, 
the contours rotate, with the degeneracy direction eventually achieving 
\beq 
3.6\,\Delta w_0+\Delta w_a={\rm constant} 
\eeq 
at $z\gg1$ (for the flat $\Lambda$CDM case of $w_0=-1$, $w_a=0$). 

\begin{figure}[!hbt]
\begin{center} 
\psfig{file=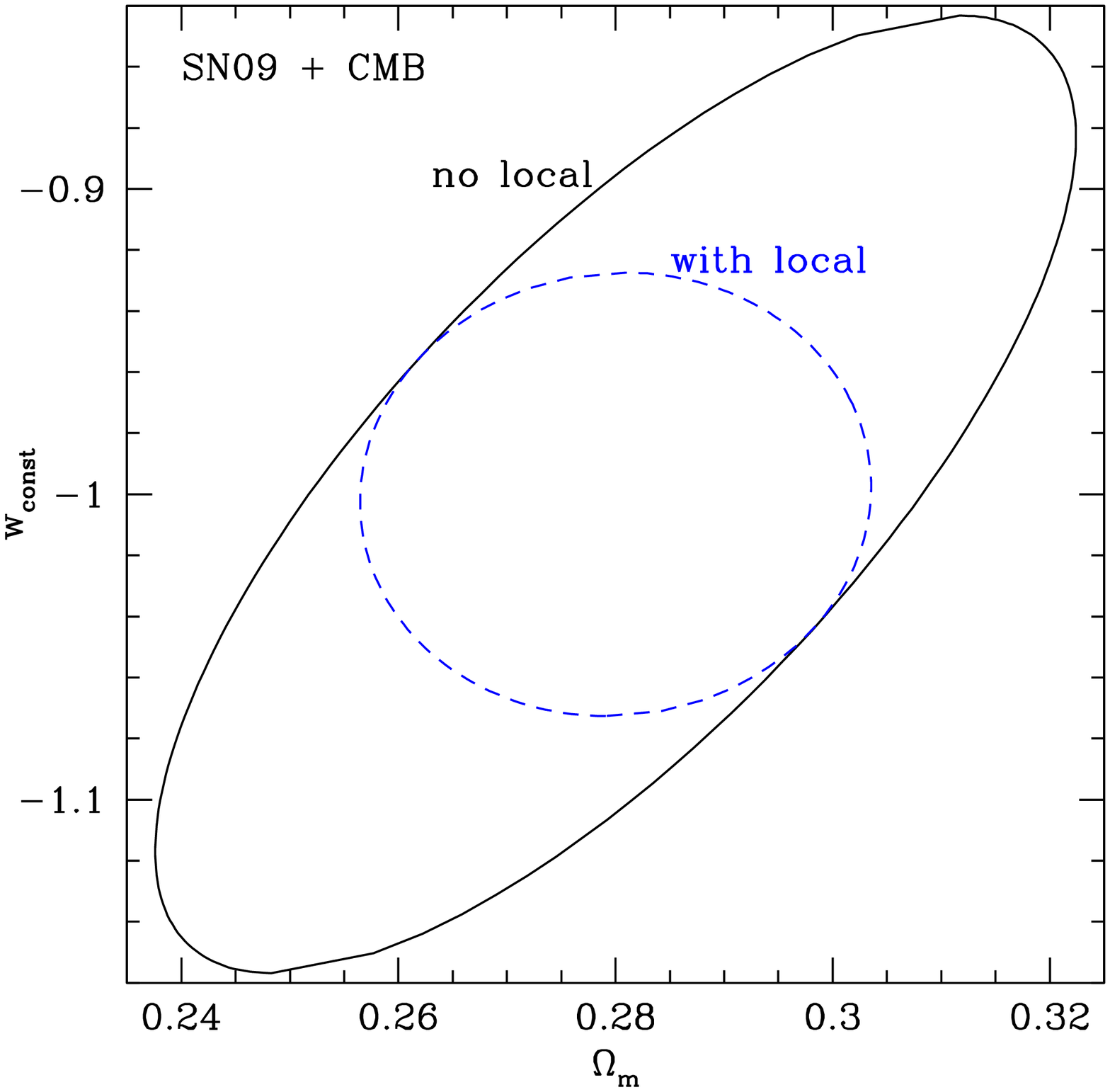,width=3.4in} 
\psfig{file=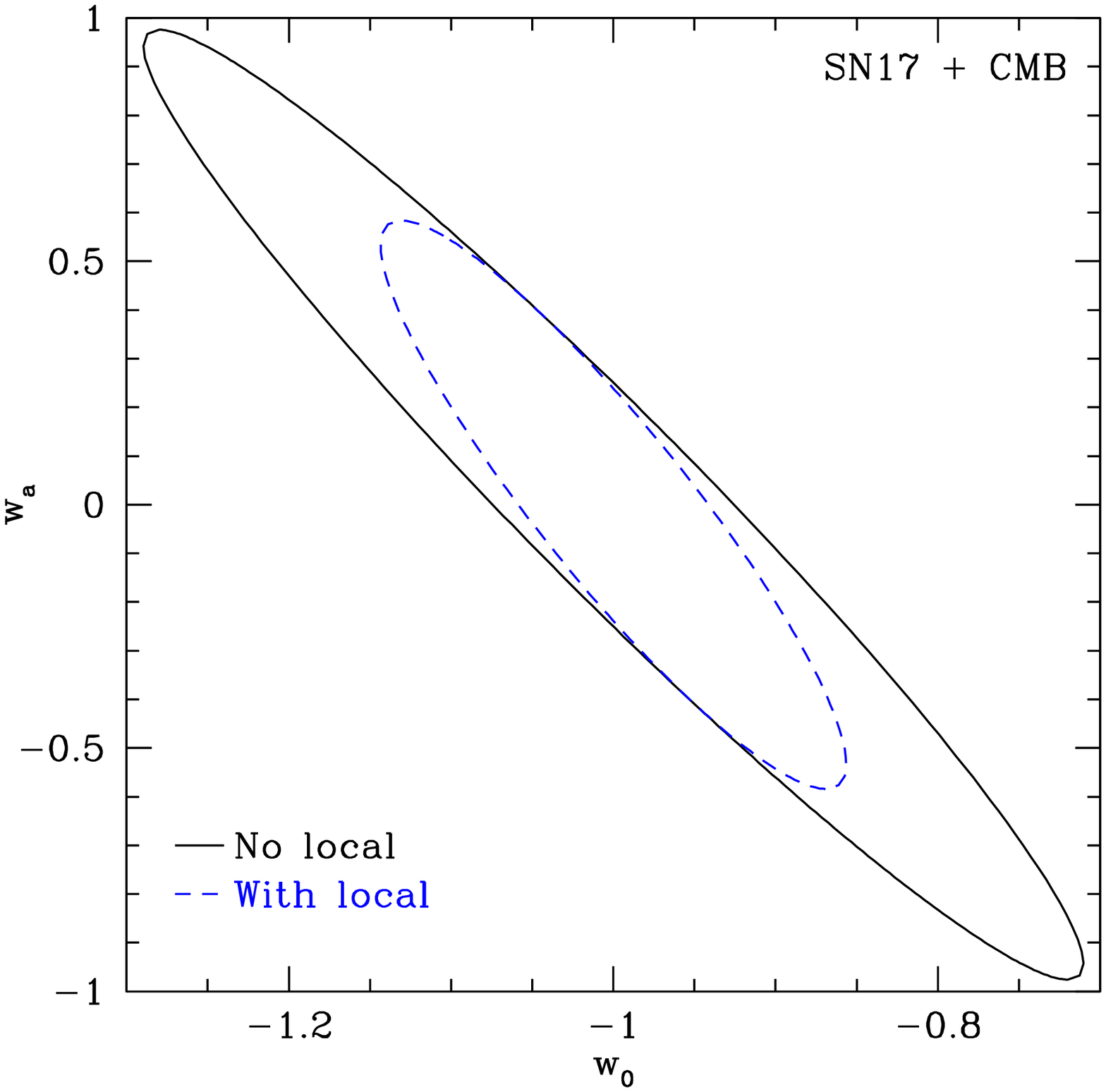,width=3.4in} 
\caption{A well controlled low redshift sample of supernova is an essential 
complement to higher redshift SN experiments.  Lack of local SN (here 300 
at $z=0.05$) diminish the cosmological leverage by a factor of two on 
both the matter density and the (assumed constant) dark energy equation 
of state for a near term SN survey to $z=0.9$ (top panel), or again by 
a factor two in the equation of state plane for a future SN survey to 
$z=1.7$ (bottom panel).  Contours show $1\sigma$ joint (68\%) confidence 
levels and systematics are included as in the text. 
}
\label{fig:snfplus} 
\end{center} 
\end{figure}

For both these reasons, local SN are essential for anchoring the Hubble 
diagram.  We illustrate their use in Fig.~\ref{fig:snfplus}, showing 
constraints on dark energy with and without local SN 
added to higher redshift data.  The local SN here comprise 300 SN at $z=0.05$ 
(we discuss the insensitivity to the redshift distribution later), with 
statistical magnitude dispersion 0.15 mag and a systematic floor of 
0.01 mag (roughly equivalent to expectations for 
the Nearby Supernova Factory \cite{snf}).  
The sample SN09 corresponds to 500 SN 
uniformly distributed between $z=0.2-0.9$ with 
dispersion 0.15 mag and a systematic of 0.04(1+z)/1.9 (roughly 
equivalent to a completed Supernova Legacy Survey \cite{snls}).  SN17 
has 2000 SN between $z=0.2-1.7$ with roughly constant numbers per 
cosmic time interval, with dispersion 0.15 mag and systematic 
0.02(1+z)/2.7 (roughly equivalent to the SN sample of the future 
Supernova/Acceleration Probe (SNAP \cite{snap})).  In all cases we 
take a flat, fiducially $\Lambda$CDM with $\om=0.28$, universe 
and include 
a CMB prior of 0.7\% on the distance to last scattering. 

Without the local SN, constraints from the current SN experiments 
will be a factor of two worse on both $\om$ and an assumed constant 
equation of state $w_{\rm const}$.  For the future SN experiment with 
tight systematics control to $z=1.7$, the leverage on the dark energy 
equation of state parameters is approximately a factor of two worse 
without local SN.  Thus it is crucial to implement a properly designed 
low redshift SN survey in order to realize the capabilities of a SN 
cosmological probe experiment.

\section{Numbers vs.\ Redshift} \label{sec:nz} 

Given the critical need for a low redshift SN experiment to complement 
higher redshift data, it is important to craft the appropriate requirements 
for the local sample to fulfill its role.  Here we investigate from a 
cosmological perspective the optimal criteria for the numbers and 
redshift distribution of the local SN to strengthen a higher redshift 
sample, within a Fisher matrix approach.  We initially treat this for 
the experiment extending to $z=1.7$ 
having the aim of understanding the dark energy properties, through its 
dynamics.  Then we consider this for the current/near term experiment 
to $z=0.9$, which can only see an averaged, or constant, view of the 
dark energy equation of state. 
\S\ref{sec:sys} returns to the issue of requirements by considering the 
role of the local sample in controlling SN systematics. 

\subsection{Numbers and leverage} \label{sec:numlever}

If one thinks of the local SN as providing a prior on the 
parameter ${\cal M}$ (which is an overstrong view as we will shortly 
see), then fixing ${\cal M}$ would require an infinite number of SN 
and the absence of any systematics floor.  However, even a perfect 
determination of ${\cal M}$ does not give an unbounded improvement on 
the other cosmological parameters.  Their estimation uncertainties 
become 
\beqa 
\sigma^2(p_j)_{{\rm fix}\ {\cal M}}&=&\sigma^2(p_j)_{{\rm fit}\ {\cal M}}- 
(C_{{\cal M}j})^2/\sigma^2({\cal M}) \\ 
&=&\sigma^2(p_j)_{{\rm fit}\ {\cal M}}\,(1-r^2), 
\eeqa 
where $C$ is the covariance matrix and $r$ is the correlation coefficient. 
For the SN17+CMB case, perfect determination of ${\cal M}$ would improve 
the estimation of $\om$, $w_0$, $w_a$ by 21\%, 70\%, 55\% respectively. 
However, as we saw from Fig.~\ref{fig:snfplus}, a local sample of 300 SN 
at $z=0.05$, with systematics, already offers substantial improvement, 
indeed 90\%, 71\%, 
73\% of the improvement possible from the unrealistic case of perfect 
determination without systematics floor. 

Furthermore, the effect of a finite sample is even closer to saturating 
the case 
of infinite numbers of local SN because local SN do not in fact constrain 
purely ${\cal M}$.  Because they are at finite distances, $z\ne0$, 
the local SN have unavoidable covariances mixing ${\cal M}$ and the 
cosmological parameters.  This will further cause a plateau in the 
numbers of local SN effective in  complementing higher redshift SN to 
impose cosmological constraints.

More and more local SN will not have continuing direct benefit for 
cosmological constraints.  To examine where the point of diminishing 
returns lies, we consider the definite case of local SN centered at 
$z_c=0.05$.  As we will see, this is close to the optimum when the 
covariance mentioned above and systematics are taken into account, 
but we emphasize that the saturation is a product of the innate 
cosmological degeneracy, independent of systematics. 

\begin{figure}[!hbt]
\begin{center} 
\psfig{file=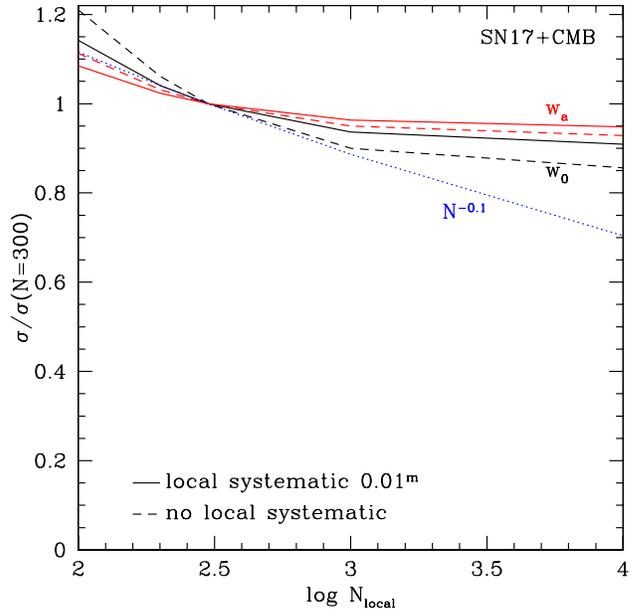,width=3.4in} 
\caption{Cosmological parameter estimation is a slow function of the 
number $N_{\rm local}$ of supernovae in the low redshift sample.  
(The curves here take $z_c=0.05$, but the behavior is similar 
over $z_c=0.03-0.15$.)   
Due to inherent cosmological degeneracy, the point of diminishing 
returns is reached near $N_{\rm local}\approx300$, whether systematics 
are present (solid curves) or not (dashed curves).  The blue dotted 
curve shows a scaling of $N^{-0.1}$ for comparison. 
}
\label{fig:snlow} 
\end{center} 
\end{figure}

Figure \ref{fig:snlow} illustrates the saturation of the cosmological 
constraints as we increase the number of local SN.  The uncertainties 
in $w_0$ and $w_a$ improve extremely little and far more slowly than 
the naive $N^{-1/2}$ behavior.  
For reference, the blue dotted line shows improvement as $N^{-0.1}$. 
We see that 300 local SN is close to optimum in strengthening the higher 
redshift SN program.  Increasing the local SN numbers beyond 300 does not 
even attain $N^{-0.1}$ improvement. 
Indeed, for $N>10^4$ the power law index is flatter than $-0.002$; 
for $N\gg 300$ the improvement over 300 SN is only 5-10\%.  
(And when we take into account below the interaction of the redshift $z_c$ 
with the cosmological degeneracy, the use of $10^4$ SN with $z_c=0.1$ still 
provides 10-12\% worse constraints than 300 SN with $z_c=0.05$.)

Overall, there is little direct cosmological use for $N>300$.  As we 
drop below 300 local SN, the degradation is faster than $N^{-0.1}$ 
for $w_0$, and for $w_a$ as we drop below 200.  Note that this trend 
remains the same whether we include systematics or not; the saturation 
is due to the innate and unavoidable cosmological degeneracy. 
(We have checked that this holds for a reduced intrinsic magnitude 
dispersion of 0.1 mag as well, though 
as the ratio of dispersion to systematic level decreases the sample 
saturates at a slightly smaller size.) 
Thus, for a homogeneous sample, 300 local SN 
is close to optimal for cosmological parameter constraints\footnote{This 
can be generalized: to match a given number of higher redshift SN, one should 
have approximately a 1:6 ratio of local:higher redshift SN.  This is somewhat 
smaller than the delta function optimization ratio 1:3 for fitting four 
parameters (see \cite{huttur}) because of the addition of CMB 
data and systematics, which spread the delta functions over redshift 
\cite{fhlt}.  Also see \S\ref{sec:snls}.}. 

\subsection{Redshift and leverage} \label{sec:redlever} 

We now examine the optimal redshift location, given the presence of the 
cosmological degeneracy.  Returning to our earlier point about covariance 
between parameters for a sample at finite redshift, $z_{\rm loc}$, 
we seek to minimize the degeneracy.  To gain an intuitive feeling for this, 
consider the local SN as determining the magnitude $m(z_{\rm loc})$ to 
precision $\sigma_{\rm loc}$.  This data can be viewed (purely 
illustratively) as putting a prior on 
\beq 
Y\equiv {\cal M}+5\log d_l(z_{\rm loc}), 
\eeq 
which is added to the Fisher information of the higher redshift SN through 
\beq 
F^{\rm prior}_{ij}=\frac{1}{2\sigma_{\rm loc}^2}\,\frac{\partial^2(Y- 
\bar Y)^2}{\partial p_i \,\partial p_j}, 
\eeq 
where $p_i$ are the cosmological parameters.  If 
$z_{\rm loc}\ll1$ then we can calculate analytically the effect of the 
prior.  

Because the luminosity distance 
\beq 
d_l(z_{\rm loc}\ll1)\approx z+(z^2/4)[1-3w_0(1-\om)], \label{eq:dloc}
\eeq 
the low redshift SN do not determine ${\cal M}$ but rather constrain 
a certain combination of ${\cal M}$, $\om$, and $w_0$.  An unavoidable 
degeneracy remains, regardless of how many local SN are measured. 
The cross terms between ${\cal M}$ and the cosmology parameters 
are proportional to $z_{\rm loc}$ and the cosmology-cosmology terms 
go as $z_{\rm loc}^2$.  To minimize the degeneracy, and hence increase 
the cosmological constraining power, we need to minimize $z_{\rm loc}$ 
(subject of course to other uncertainty sources such as peculiar 
velocities, which we address below).  

The illustrative analytic expression of 
Eq.~(\ref{eq:dloc}) agrees extremely well with the numerical calculations 
(which we always use), pointing up the inherent 
cosmological degeneracy that leads to the saturation in the use of large 
numbers of local SN for cosmological leverage and the preference for 
lower redshift.  
Thus, local SN at $z=0.05$ strengthen the higher redshift program much 
more than SN at $z=0.1$.  For example, with 300 local SN, the degradation 
in constraints if $z_{\rm loc}=0.1$ rather than 0.05 is 24\%, 16\% 
for $w_0$, $w_a$ (and goes roughly linearly with deviation of $z_{\rm loc}$ 
from 0.05).  As already seen, even large numbers of SN at $z_{\rm loc}=0.1$ 
cannot overcome this disadvantage.  

As expected, the cosmological degeneracy imposes a monotonic optimization, 
pushing $z_{\rm loc}\to0$.  This, however, is impractical for a realistic 
experiment.  Two factors 
dominate in working against very low redshift: the smaller volume available 
and hence fewer SN, and uncertainties contributed to 
the distance determination by random and 
coherent peculiar velocities.  These raise the very low redshift end of 
the optimization curve, creating a minimum at a finite redshift. 

To take into account the volume effect, we realize that an experiment 
with fixed survey time, centered at $z_c$, can amass a number $N$ of well 
characterized SN (i.e.\ not just discovered, but followed up with 
spectroscopy), where 
\beq 
N\sim \int_{\rm bin} dz\,z^2 A(z), 
\eeq 
where $A(z)$ gives the dependence on the redshift depth of the 
solid angle that can be covered in the survey time.  This accounts for the 
increased amount of 
time required to observe fainter SN.  We approximate $A(z)\sim z^{-\gamma}$, 
which includes the cases of sky noise domination ($\gamma=4$) and source 
noise domination ($\gamma=2$), and should provide a reasonable fit between 
these two limits.  Both $N$ and $A$ are normalized to the $z_c=0.05$ 
case of 300 well characterized SN in 20000 deg$^2$.  We limit the sky 
area to a maximum of 30000 deg$^2$ and take the total redshift bin width 
to be 0.05, centered at $z_c$.  We have checked that the exact distribution 
of SN around the central redshift has very weak influence, of order 1\% in 
the parameter constraints, i.e.\ it does not matter if within the bin the 
SN are taken to be all at the bin center, uniformly spread, or scaled 
with the local volume element. 

Effects due to peculiar velocities of galaxies in which SN reside consist 
of a statistical error due to random velocities (which we take to be 
300 km/s) and a systematic error due to bulk motions.  The latter is 
treated following the formalism of \cite{huigreene,cooraycaldwell} and 
we approximate their results by an irreducible error across the local 
redshift bin of $\sigma^2_{\rm vsys}\sim A^{\alpha}z_c^{-\beta}$, added 
to the SN random variance.  We find a good fit with 
\beq 
\sigma_{\rm vsys}=0.0077^m\left(\frac{A}{20000\,{\rm deg}^2}\right)^{-1/4} 
\left(\frac{z}{0.065}\right)^{-3/2}. \label{eq:vsys}
\eeq 

\begin{figure}[!hbt]
\begin{center} 
\psfig{file=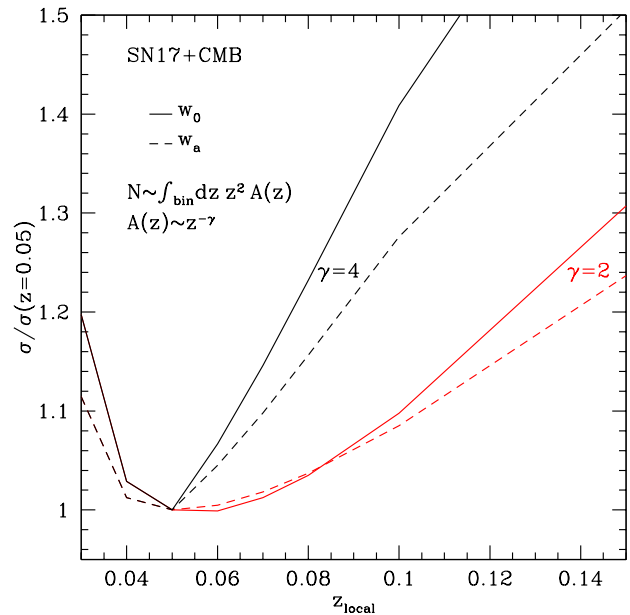,width=3.4in} 
\caption{The interaction of cosmological degeneracies, inherent 
even at low redshift, velocity flow systematics, and observational 
considerations creates a optimum redshift for the low redshift sample 
of supernova serving to anchor the Hubble diagram.  Here we account 
for all these effects, with the number of observed supernovae scaling 
with redshift depth according to the available volume, but also 
the more limited solid angle $A(z)$ that can be covered in fixed survey 
time (due to fainter source magnitudes).  The optimum central redshift 
of the low redshift supernovae is $z\approx0.05$. 
}
\label{fig:snlowloc} 
\end{center} 
\end{figure}

Putting all this together, Fig.~\ref{fig:snlowloc} shows the realistic 
dependence of the cosmological constraints on $z_{loc}$.  There is now 
a clear optimum location for the local SN sample, at $z_c=0.05-0.06$, to 
maximize the science return of the SN experiments.  Recall that the 
true area scaling will lie between the $\gamma=2$ and $\gamma=4$ cases, 
with $\gamma=2$ (source noise domination) holding for more nearby SN 
observed near peak brightness and $\gamma=4$ (sky noise domination) 
holding for more distant SN or observations away from peak brightness. 
At very low redshift, the ceiling on the area makes the results independent 
of $\gamma$.

Technically, because of the volume weighting, the bin center is not 
the same as the mean redshift.  For example, for a sample spanning 
$z=0.03-0.08$ (such as the Nearby Supernova Factory), the weighted 
mean is $\langle z\rangle=0.062$.  Interestingly, the standard 
expression for SN systematics for the SNAP-like sample, $0.02(1+z)/2.7$ 
(see \S\ref{sec:cos}), predicts 0.0079$^m$ here, hence essentially 
already providing a good approximation to the local sample systematics 
expressed by Eq.~(\ref{eq:vsys}). 

The conclusion about the optimum $z_{\rm loc}$ remains robust in the 
presence of a further systematic involving a magnitude offset between 
the local 
sample and the higher redshift sample, due to calibration for example. 
Such a step or offset has long been recognized as a possible observational 
feature, and treated in both its random and coherent aspects in various 
analyses such as \cite{klmm,linmiq,linbias}.  Considering offsets limited 
by Gaussian priors of 0.01 or 0.02 mag, we find that this does not change 
the shape of the curves in Fig.~\ref{fig:snlowloc} nor the location of 
the optimum at $z_{\rm loc}\approx 0.05$. 

\subsection{Matching current supernovae surveys} \label{sec:snls}

It is of considerable interest to consider as well the optimization of 
the nearby sample for complementing ongoing and near 
term SN experiments that provide data sets similar to the SN09 case of 
\S\ref{sec:cos}, i.e.\ extending to $z=0.9$.  Recall that 
Fig.~\ref{fig:snfplus} showed that 
the local sample played a crucial role here too.  The key cosmological 
effects and formalism remain the same as in 
\S\ref{sec:numlever}-\ref{sec:redlever}. 

\begin{figure}[!hbt]
\begin{center} 
\psfig{file=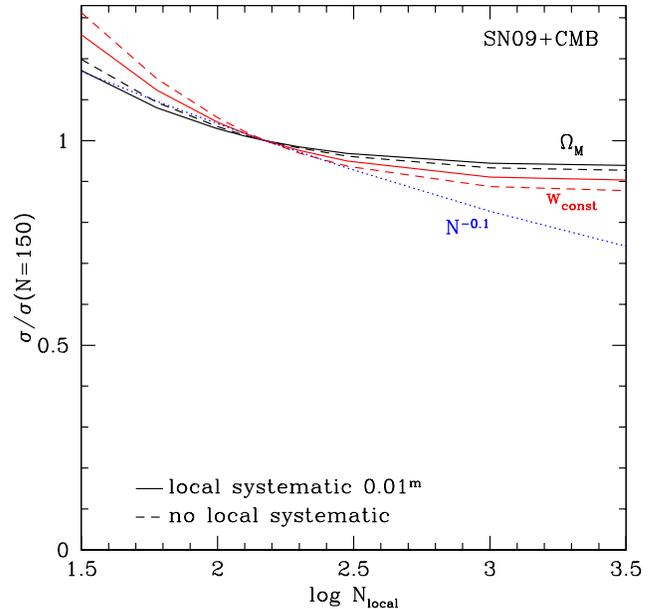,width=3.4in} 
\caption{As Fig.~\ref{fig:snlow} but using the higher redshift supernova 
sample SN09 similar to ongoing surveys (in particular extending to 
$z_{\rm hi}=0.9$).  Diminishing returns in cosmological leverage -- now 
on the matter density $\Omega_M$ and constant dark energy equation of 
state $w_{\rm const}$ -- are 
reached near $N_{\rm local}\approx125-150$. 
}
\label{fig:snlow9} 
\end{center} 
\end{figure}

Figure \ref{fig:snlow9} again illustrates the saturation of cosmological 
leverage as the number of SN increases.  Here we see that 125-150 local SN 
prove sufficient to serve as the low redshift anchor for the Hubble 
diagram.  The ratio of local:higher redshift SN is now 1:4.  The reduced 
ratio makes sense since for this case we fit only three parameters -- 
${\cal M}$, $\om$, $w_{\rm const}$.  As in the footnote in 
\S\ref{sec:numlever}, the idealized optimum would be SN distributed in 
$P$ delta functions in redshift, where $P$ is the number of fit parameters, 
with one of them at $z\approx0$. 
As mentioned there, the inclusion of systematics and priors spreads out 
the delta functions, reducing the idealized ratio by about a factor of 
two.  Because fewer local SN are involved, diminished intrinsic 
magnitude dispersion will now have a larger effect; we find that for 
0.1 mag dispersion the saturation occurs near 60 local SN.

\begin{figure}[!hbt]
\begin{center} 
\psfig{file=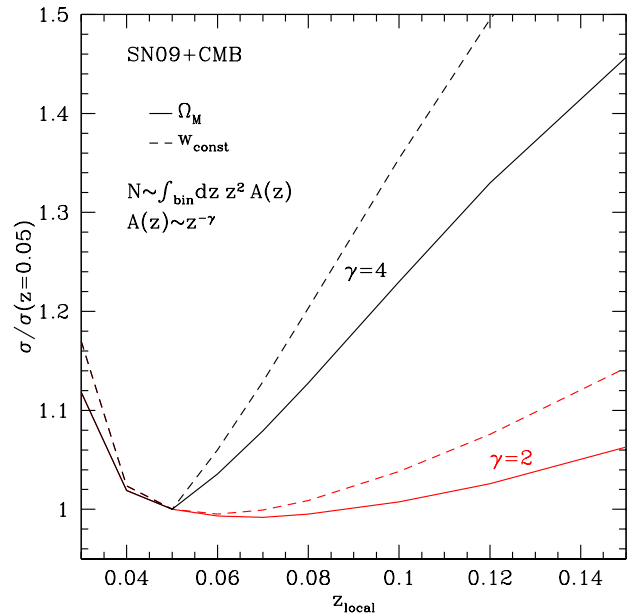,width=3.4in} 
\caption{As Fig.~\ref{fig:snlowloc} but using the higher redshift supernova 
sample SN09 similar to ongoing surveys (in particular extending to 
$z_{\rm hi}=0.9$).  The optimum central redshift of the low redshift 
supernovae is $z\approx0.05$. 
}
\label{fig:snlowloc9} 
\end{center} 
\end{figure}

The redshift optimization for the local SN remains at $z\approx0.05$, 
as illustrated in Fig.~\ref{fig:snlowloc9}.  The key cosmological 
influences of innate distance degeneracy and velocity flow systematics 
are intrinsic to the local sample and so the optimum central redshift 
for the local SN is insensitive to the higher redshift sample: we again 
find the optimum is $z_{\rm loc}\approx0.05$.  As before, this remains 
robust under introduction of a magnitude offset between the local and 
higher redshift sample.

\section{Subsampling} \label{sec:sys}

The second purpose for the local SN program involves identification of 
SN systematics, characteristics that would break the homogeneity of the 
sample.  Control of such systematics 
is frequently discussed in terms of subsamples and like vs.\ 
like comparison (see, e.g., \cite{coping}).  
Let us estimate the number of local SN required for such subsampling.  
An important point to be made is that a difference that 
makes no difference is no difference -- and furthermore a difference that 
cannot be detected is no difference.  That is, defining 
a subset with certain lightcurve and spectral properties distinct from 
another subset is of limited impact (for cosmology) if they do not differ in 
their corrected peak magnitudes.  Furthermore, a difference in subsample 
mean magnitudes 
smaller than the uncertainty from, say, calibration is likewise moot. 

\subsection{Population drift} \label{sec:drift}

An important issue to begin with, then, is how large a magnitude difference 
has a substantial cosmological effect.  We call subsets, defined on the 
basis of empirical differences in the lightcurves and spectra, subclasses 
if they differ in corrected peak magnitude.  Subclasses will combine 
together to determine the mean magnitude at a given redshift; as long as 
the proportions among subclasses remain independent of redshift there is no 
cosmological consequence.  That is, it is not the presence of subclasses 
per se, but the population drift and hence weighting in the mean magnitude 
at each redshift that is important.  

Consider the most extreme subclass, subclass 2, lying furthest from the 
mean magnitude of the other subclasses, collectively subclass 1, in a 
sort of jackknife test.  
Suppose its absolute magnitude deviates from the 
mean absolute magnitude by $\Delta M_{21}$ and it comprises a fraction 
$f_2(z)$ of the total number at redshift $z$.  Then the evolution in the mean 
magnitude relative to the local sample is 
\beq 
\Delta M(z)=\Delta M_{21}\,[f_2(z)-f_2(0)]. 
\eeq 

Employing the Fisher bias formalism we can calculate the effect of 
an evolution $\Delta M(z)$ in biasing cosmological parameter extraction. 
By requiring that the bias on a parameter $\delta p_i<0.46\,  
\sigma(p_i)$, we ensure that the risk $[\sigma^2(p_i)+ \delta p_i^2]^{1/2}$ 
degrades the standard deviation by less than 10\%.  We can then ask 
what is the largest acceptable $\Delta M_{21}$ for a given population 
drift function $F(z)=f_2(z)-f_2(0)$.  For each of three forms: 
$F(z)\sim z$, $1-a$, or $z^2$, 
we find the requirement $\Delta M(z=1.7)<0.016$, for the SN17 case 
(see the end of this subsection for the SN09 case).  This arises mostly 
from preventing bias in $\om$; for the $z^2$ drift the constraints from 
$w_0$ and $w_a$ are also comparable, but for the $z$ behavior the next 
tightest constraint is 0.026 from $w_a$, and for the $1-a$ case it is 
0.066 from $w_0$.  So 0.016 mag is a fairly conservative criterion. 

The strongest constraint on the observable $\Delta M_{21}$ comes from the 
most extreme case, where $F(z=1.7)=1$.  So we have the requirement of 
recognizing a subclass with $\Delta M_{21}>0.016$ mag.  This is 
fortunate -- if we have a calibration systematic of 0.005 
mag then this represents  a healthy $3\sigma$ effect.  Had the 
requirement been to distinguish subclasses differing by less than 0.005 
we would have been in trouble.  We can now flow this down to a requirement 
on the local SN. 

Note that what concerns us is not the evolution as such, but the 
uncertainty in the 
evolution.  If the population drift is rapid, $f_2(z=1.7)\gg f_2(0)$, 
then the magnitude evolution will be insensitive to $f_2(0)$.  The 
uncertainty will arise from the precision on knowing $M_1$ (the 
absolute magnitude of the main population) and $M_2$, and any correlation 
(e.g.\ from calibration).  We learn about $M_1$ and $M_2$ from studying 
the local sample, where we know the distance.  If the dispersions of 
subsamples 1 and 2 are similar, but population 2 is much rarer, $N_2\ll N_1$, 
then the uncertainty on $\Delta M_{21}$ is dominated by the number of 
local SN in subclass 2.  

To determine the mean magnitude of 
a subclass to 0.016, in the presence of measurement dispersion of 0.15 
mag, we need $N_2=(0.15/0.016)^2=88$ local SN of this subclass. 

The results when considering the SN09 case are similar.  The requirement 
on the magnitude drift, or subclass deviation, becomes 
$\Delta M(z=0.9)=0.019-0.025$; this implies a need for 36-62 local SN of 
this subclass. 

\subsection{Improved standardization} \label{sec:disp}

One instance in which a subset that is not a subclass (i.e.\ does not 
differ in its mean peak magnitude) can be useful is when the subset 
exhibits decreased dispersion about the mean, that is, when it is a 
more standardized candle.  Identification of such subsets within the 
local sample is of interest, and is a positive effect as opposed to the bias 
caused by drifting subclasses.  (However note that explaining the 
full sample dispersion of, say, 0.15 mag by multiple subclasses of 
smaller dispersion requires the means of these subclasses to differ.) 

The effect of reduced dispersion, in particular of just a subset, 
may not be dramatic.  If 1/3 of the SN have dispersion 0.05 mag rather 
than 0.15 mag, this only reduces the overall dispersion to 0.126 mag. 
If the overall dispersion is reduced to 0.1 mag from 0.15 mag, then the 
equation of state parameters $w_0$, $w_a$ improve by 15\% (for the 
canonical survey characteristics).  Reduced dispersion to 0.1 does not 
much affect 
the bias level: the requirement is $\Delta M(z=1.7)<0.015$ for 
$F(z)\sim z^2$ and 0.016 for $F(z)\sim z$, $1-a$.  However this does mean 
that the subclass magnitude can be determined with $(0.1/0.015)^2=44$ SN, 
half the number needed previously\footnote{For the SN09 case, the 
requirement is $\Delta M(z=0.9)<0.017-0.024$, so the number needed is 
17-35, again roughly half the 0.15 mag dispersion result.}.  Thus 
subsets with reduced dispersion can 
reduce the number of local SN needed to guard against subclass bias, 
and the numbers of local SN we quote below are therefore conservative. 

\subsection{Two stage program} \label{sec:stage}

To harvest a subset of size $N_2$, which represents a fraction $f_2(0)$ 
of the full population, 
requires a local sample totaling $N_2/f_2(0)$.  Thus, if the most extreme 
subclass is also the rarest, we face a challenge.  Suppose subclass 2 
represented only 5\% of SN at $z=0$, but close to 100\% at $z=1.7$.  Then 
we would need $88/0.05\approx1800$ local SN.  

To ameliorate the factor 20 increase in numbers, one could possibly 
design the low redshift SN program to concentrate on the extreme examples. 
One approach would be first to obtain the cosmological leverage number of 
300 SN, then concentrate in a second survey stage on the most extreme 
subclasses (ones deviating from the mean by more than 0.016 
mag) and ensure samples of $\sim$100 SN for them.  This would involve much 
more searching time, but not a large increase in follow up -- if the 
subclass could be recognized without the full set of measurements.  If 
successful, such pruning could keep the total sample to 400-900 SN. 

We emphasize that what is important in determining the number of local 
SN is not the number of subsets, or even subclasses, but the number of 
extreme ($\Delta M_{21}>0.016$), differentially drifting subclasses.  
If there is more than one extreme subclass, then this decreases $f_2(z=1.7)$, 
say, below unity, and hence loosens the requirement on $\Delta M_{21}$, 
changing the definition of extreme.   
Thus it is not unlikely that only one or two extreme subclasses exist in 
this sense and hence require a sample of only 100-200 Stage 2 local SN 
in addition to the Stage 1 foundation program of 300 local SN. 

The two stage approach also has the virtue that the initial 300 local SN 
are likely to prove a sufficient data set by itself for the ongoing/near term 
higher redshift SN program, even with extreme, differentially drifting 
subclasses. 

\section{Conclusions} \label{sec:further} 

Local supernovae serve critical roles 
for maximal science return from supernova cosmology experiments -- 
anchoring the Hubble diagram and identifying possible systematics -- 
and failure to supply these data can cost a factor of 2 in 
cosmology parameter determination. 
General considerations of cosmology in the form of 
parameter degeneracies, systematics, 
and biases impose reasonable and somewhat conservative criteria for a 
low redshift SN program to match and strengthen a comprehensive, 
next generation higher SN program: 300-900 SN centered at 
$z\approx0.05$.  We have tested the robustness of this conclusion by 
including coherent and random velocity systematics, magnitude offsets, 
and different intrinsic dispersions.  

We propose here a two stage approach to the local sample, with the 
foundational set composed of 300 well characterized local SN.  This data 
then determines the necessary size for an extension, or second stage, 
which the arguments presented here suggest may require 100-600 additional 
SN to control systematic biases.  The first stage would likely supply the 
necessary local set to fully match ongoing and near term SN surveys, which we 
concluded also are optimized by a local data set centered at $z\approx0.05$. 

Though we used straightforward cosmological considerations, for more 
detail we could also seek future guidance from SN theory on questions 
such as maximum expected magnitude evolution, population drift predictions, 
and characteristics for subclasses, and from calibration and measurement 
error models and simulations.  This detail may be unnecessary and is 
model dependent.  The empirical foundations -- from well characterized 
(i.e.\ spectroscopic time series measured) SN, not just more SN -- 
 will define the relation between subsets 
and subclasses, and the population distribution (e.g.\ local percentage 
of low metallicity SN, differing in the mean magnitude by $X$ mag).  
The data will be the final arbiter of the sample size 
required\footnote{Note that much valuable information will be gained 
from surveys, even without full spectroscopic followup, including 
Carnegie Supernova Project \cite{csp}, CfA Nearby SNe \cite{cfa}, 
CTI-II \cite{cti}, Essence \cite{essence}, KAIT \cite{kait}, 
Nearby Supernova Factory, 
SDSS II \cite{sdss2}, SkyMapper \cite{skymap}, Supernova Legacy Survey, 
and Texas Supernova Search \cite{quimby}.}.  

To achieve understanding of the physics behind the acceleration of 
the universe, we need a well designed SN program at $z<0.1$ just as 
we need a comprehensive SN experiment extending to $z>1.5$. 
Implementing the criteria derived here should ensure that we reach out 
into the universe with a firm foundation beneath our feet.

\section*{Acknowledgments} 

I thank Greg Aldering for discussion of many aspects of nearby SN 
observing, especially the limiting scalings of survey area with redshift, 
and Stu Loken and Mark Strovink for probing questions. 
This work has been supported in part by the Director, Office of Science,
Department of Energy under grant DE-AC02-05CH11231.


\begin{thebibliography}{99}

\bibitem{perl99}
S. Perlmutter et al. 1999, ApJ 517, 565

\bibitem{riess98}
A.G. Riess et al. 1998, AJ 116, 1009 

\bibitem{snls}
  Supernova Legacy Survey: http://cfht.hawaii.edu/SNLS\\ 
  P. Astier et al. 2006, A\&A 447, 31

\bibitem{knop03}
R.A. Knop et al.\ 2003, ApJ 598, 102 

\bibitem{riess04}
A.G. Riess et al.\ 2004, ApJ 607, 665

\bibitem{lh03}
E.V. Linder \& D. Huterer, D.\ 2003, Phys.\ Rev.\ D 67, 081303

\bibitem{fhlt} 
  J.A.\ Frieman, D.\ Huterer, E.V.\ Linder and M.S.\ Turner 2003, 
  Phys.\ Rev.\ D, {\bf 67}, 083505 

\bibitem{snf}
  Nearby Supernova Factory: http://snfactory.lbl.gov ; \\ 
  M. Wood-Vasey et al. 2004, New Astron. Rev. 48, 637

\bibitem{snap}
  SNAP -- http://snap.lbl.gov ; \\ 
  G.\ Aldering et al.\ 2004, astro-ph/0405232

\bibitem{huttur} 
  D. Huterer \& M.S.\ Turner 2001, Phys. Rev. D 64, 123527 

\bibitem{huigreene}
L. Hui \& P.B. Greene 2006, Phys. Rev. D 73, 123526

\bibitem{cooraycaldwell}
A. Cooray \& R.R. Caldwell 2006, Phys. Rev. D 73, 103002

\bibitem{klmm}
A.G. Kim, E.V. Linder, R. Miquel, N. Mostek 2004, MNRAS 347, 909

\bibitem{linmiq}
E.V. Linder \& R. Miquel 2004, Phys. Rev. D 70, 123516 

\bibitem{linbias}
  E.V.\ Linder 2006, Astropart. Phys. 26, 102 

\bibitem{coping}
  D. Branch, S. Perlmutter, E. Baron, P. Nugent 2001, in Resource 
  Book on Dark Energy, ed. E.V. Linder [astro-ph/0109070] 

\bibitem{csp} 
http://www.ociw.edu/csp ; \\ 
W.L.\ Freedman et al.\ 2005, ASP Conf. Proc. 339, 50 
[astro-ph/0411176] 

\bibitem{cfa}
http://cfa-www.harvard.edu/oir/Research/supernova

\bibitem{cti}
  CCD/Transit Instrument II: J.T.\ McGraw et al.\ 2006, SPIE 6267, 126 

\bibitem{essence}
    Essence: http://www.ctio.noao.edu/wproject ; \\ 
J.\ Sollerman et al. 2005, astro-ph/0510026

\bibitem{kait} 
  http://astro.berkeley.edu/~bait/kait.html

\bibitem{sdss2}
http://sdssdp47.fnal.gov/sdsssn/sdsssn.html

\bibitem{skymap} 
http://www.mso.anu.edu.au/skymapper 

\bibitem{quimby} 
http://grad40.as.utexas.edu/$\sim$quimby/tss  


\end{thebibliography}
\end{document}